\documentstyle[11pt,aaspp4,tighten]{article}

\def\today{\rightline{\ifcase\month\or
        January\or February\or March\or April\or May\or June\or
        July\or August\or September\or October\or November\or December\fi
        \space\number\day, \number\year}}
\def\etal{{\it et al.}\ }

\def\s-1{{\rm\,s^{-1}}}

\def\spose#1{\hbox to 0pt{#1\hss}}
\def\C3H2{{\rm\,\rm C_3H_2}}
\def\NH3{{\rm\,\rm NH_3}}
\def\HOCO+{{\rm\,\rm HOCO^+}}
\def\lta{\mathrel{\spose{\lower 3pt\hbox{$\mathchar"218$}}
     \raise 2.0pt\hbox{$\mathchar"13C$}}}
\def\gta{\mathrel{\spose{\lower 3pt\hbox{$\mathchar"218$}}
     \raise 2.0pt\hbox{$\mathchar"13E$}}}

\begin{document}

\font\twelvei = cmmi10 scaled\magstep1 
       \font\teni = cmmi10 \font\seveni = cmmi7
\font\mbf = cmmib10 scaled\magstep1
       \font\mbfs = cmmib10 \font\mbfss = cmmib10 scaled 833
\font\msybf = cmbsy10 scaled\magstep1
       \font\msybfs = cmbsy10 \font\msybfss = cmbsy10 scaled 833
\textfont1 = \twelvei
       \scriptfont1 = \twelvei \scriptscriptfont1 = \teni
       \def\mit{\fam1 }
\textfont9 = \mbf
       \scriptfont9 = \mbfs \scriptscriptfont9 = \mbfss
       \def\bmit{\fam9 }
\textfont10 = \msybf
       \scriptfont10 = \msybfs \scriptscriptfont10 = \msybfss
       \def\bmsy{\fam10 }

\def\etal{{\it et al.~}}
\def\eg{{\it e.g.}}
\def\ie{{\it i.e.}}
\def\lsim{\raise0.3ex\hbox{$<$}\kern-0.75em{\lower0.65ex\hbox{$\sim$}}} 
\def\gsim{\raise0.3ex\hbox{$>$}\kern-0.75em{\lower0.65ex\hbox{$\sim$}}} 
\today
\title{ A SPECTRAL LINE SURVEY FROM 138.3 TO 150.7 GHZ TOWARD ORION-KL 
}

\author{Chang Won Lee, Se-Hyung Cho, and Sung-Min Lee}

\vskip 0.2in
\affil{Taeduk Radio Astronomy Observatory, Korea Astronomy Observatory,}
\affil {San 36-1, Hwaam-dong, Yusung-gu, Taejon 305-348, Korea}
 
\affil{E-mail: cwl@trao.re.kr, cho@kao.re.kr, smlee@trao.re.kr}

\vskip 1in
\begin{abstract}
We present the results of a spectral line survey 
from 138.3 to 150.7 GHz toward Orion-KL. 
The observations were made using the 14 m radio telescope of 
Taeduk Radio Astronomy Observatory.
Typical system temperatures were between 500 and 700 K, with
the sensitivity between $0.02 - 0.06$ K in units of $\rm T_A^*$.
 
A total of 149 line spectra are detected in this survey.
Fifty lines have been previously reported, 
however we find 99  new detections.
Among these new lines, 32 are  `unidentified', while
67 are from molecular transitions with known identifications.
There is no detection of H or He recombination lines.
The identified spectra are from a total of 16 molecular species and
their isotopic variants. 
In the range from 138.3 to 150.7 GHz, the strongest spectral line is  
the J=$3-2$ transition of CS molecule, followed by
transitions of  
the $\rm H_2CO$, $\rm CH_3OH$, $\rm CH_3CN$, and $\rm SO_2$.
Spectral lines from the large organic molecules 
such as $\rm CH_3OH$, $\rm CH_3OCH_3$, $\rm HCOOCH_3$,
$\rm C_2H_5CN$ and $\rm CH_3CN$ are prominent; 
with 80 \% of the identified lines arising from transitions of these
molecules.
The rotational temperatures and column densities are derived using the
standard rotation diagram analysis for 
$\rm CH_3OH$ ($\rm ^{13}CH_3OH$), $\rm HCOOCH_3$, 
$\rm CH_3CN$ and $\rm SO_2$ with $\rm 10\sim 270~K$ and 
$\rm 0.2\sim 20\times 10^{15}~cm^{-2}$.
These estimates are fairly comparable to the values for 
the same molecule in other frequency regions by other studies.

\end{abstract}


\clearpage

\section{Introduction}

Orion-KL is a unique source for 
study of numerous astro-chemical aspects associated 
with massive star formation.  
Spectral line surveys can provide an unbiased
view on molecular constituents of such regions by fully probing various spectral lines in 
wide frequency bands, and also are necessary  
to understand physical and chemical processes in 
a massive star-forming region.

Toward Orion-KL there have been numerous molecular line surveys in 
the spectral ranges of $70
- 360$ GHz and $607 - 900$ GHz by other researchers (see Table 1).
However, there are no spectral line surveys in the 2 mm band 
except for $150 - 160$ GHz by Ziurys \& McGonagle (1993, hereafter ZM93). 
ZM93 have detected over 180 spectral lines including 
about 45 unidentified features, 
and so concluded that Orion-KL is also very rich region for molecular lines 
in 2 mm band. 
We have been carrying out a spectral line survey
toward Orion-KL to cover the entire `$\it unexplored$' 2 mm spectral band.
This paper presents our observational results of the 
survey from 138.3 to 150.7 GHz 
using the 14 m telescope of Taeduk Radio Astronomy Observatory (TRAO).
 
\section{Observations}

The survey used the TRAO 14 m telescope during 1999 April $-$ May.
The HPBW and beam efficiency of the telescope at 146 GHz are about
46$''$ and 39$\%$, respectively (Park et al. 1997). 
A 100/150 GHz dual channel SIS receiver which TRAO has developed (Park et al. 1999a,b) was used 
for the survey. A single sideband filter
was employed to reject image spectra from an image 
sideband. 
The rejection ratio of the image side band is over 20 dB.
Two filter banks of 256 channel of 1 MHz resolution were used in serial 
mode to get about 500 MHz band coverage during each observation.
Observations were conducted with 25 different frequency settings 
to cover the entire frequency range between 138.3 and 150.7 GHz.
Pointing of the telescope and focus of the secondary mirror were checked every two hours
using Orion-KL itself in the SiO J$=2-1,~v=1$ maser line.  
The pointing accuracy is better than $10''$, with
typical system temperature 
between 500 $-$ 700 K and a sensitivity of 0.02 $-$ 0.06 K in units of
$\rm T_A^*$.
The data were taken in position switching  mode using 
an emission free reference position of $30'$ azimuth off the source.
Spectral line intensities were calibrated, corrected for atmospheric losses
using the chopper wheel method, and expressed in $\rm T_A^*$ scale. 
The position of Orion-KL used for the observations is
($\alpha, \delta)_{1950.0}=(5^h~32^m~ 46^s.9,~-5^\circ~24'~23'')$ 

\section{Results and Discussion}

A total of 149 spectral lines are detected with
a complete spectrum shown in Fig. 1.
Detailed identification of the molecular emission lines, LTE rotation diagram analysis of 
some molecular species, other interesting lines, and spatial origin of the molecular species
are described in this section.
 
\subsection{Identification of detected lines}
The rest frequency scale for identification of the detected molecular emission lines
is obtained by  assuming 
$\rm V_{LSR} = 8.4~km~s^{-1}$. This value  
is determined from a Gaussian fit to the CS J$=2-1$ spectrum.
This can be somewhat arbitrary as the reference value,
because Orion-KL is known to consist of a number of 
kinematically different components with sizes of a few $10''$, such as
extended ridge, hot core, compact ridge, and plateau 
(Sutton et al. 1995). 
The existence of these different velocity components may yield 
an uncertainty in the derived frequency of
$\sim$ 1 MHz or slightly larger.

We used two catalogs for line identification: the Lovas catalog (Lovas 1992)
which is a collection of the molecular line frequencies
determined from astronomical observations, and  
the JPL catalog (Pickett et al. 1998), a collection of
theoretical calculations of molecular lines.
In addition, catalogs by Anderson et al. (1992), Xu \& Lovas (1997), and 
Tsunekawa et al. (2000, in private communication),
and the Kawaguchi catalog (Kawaguchi 1999, in private communication)
are complementarily used  for the line identification. 

If a detected line matches the frequency of any identified line
in Lovas catalog 
to within $\sim \pm$ 1 MHz, 
it is regarded as previously `observed'.    
However, if a detected line is not in the Lovas catalog, but 
in the list of one of the other five catalogs, 
it is considered as a `new' detection.
This case requires that any other transitions of the species 
responsible for the `new' line  should have been previously detected from 
other astronomical observations.
We mark `NDT' in the 6th column of 
Table 2 for the `newly' detected transition. 

 If a detected line is not listed in any catalog, we 
designate it as an unidentified or ``U'' line.  Occasionally
a high excitation transition of a known species matches 
the frequency of a detected line, although no lower
excitation transitions of the species have been detected.
These matches are regarded as coincidences and marked as unidentified.

With a detection limit of $\sim 3\sigma$,  
a total of 149 molecular line spectra 
are detected.
The detailed identifications of the spectra are given in Table 2.
The frequency in the first column is the value given from 
the catalogs provided in the 5th column. 
For the `U' lines, 
we give the frequencies determined from our observations. 
Molecular species, their transitions, intensity in $\rm T_A^*$, and
special comments for the spectra are given in the 2nd, 3rd, 4th, 
and 6th column of the Table 2.
  
Fifty of 149 spectral lines are found to  have been previously
identified from other studies, 
while there are 99  lines which represent the first detections towards
an astronomical source.
Sixty-seven of the new 99  lines are transitions predicted 
from theoretical calculations or detected from laboratory measurements.
However, 32 detected lines remain ``unidentified''. 
We display all scans with identifications of the spectra in Fig. 2.

The spectral region between $150000$ MHz $\sim 150760$ MHz overlapped with 
the survey region by ZM93.
ZM93 detect $\sim$19 spectral lines in this region while
we detect only 12 lines.  The sensitivity  between the two surveys
is similar and the antenna temperatures for the strong lines 
(e.g., transitions of $\rm CH_3OH$, $\rm SO_2$, and $\rm H_2CO$) 
for both observations are comparable. Moreover, the 2mm beam efficiency
of the two telescopes 
is about the same ($\sim0.4$).
The identifications for strong eight lines  are the same 
between two studies, but for weak lines, different identifications 
are given. 
We especially note two transitions of $\rm CH_3OCH_3$ at 150163.1 MHz
and 150467.3 MHz. The former 150163.1 MHz line is considered to 
be `unidentified' by ZM93 while the latter 150467.3 MHz line is not on their list.       
ZM93 assign the spectra at 150415.1 MHz and 150449.2 MHz as transitions 
of $\rm CH_3OH$ and $\rm HCOOCH_3$, respectively.
However, the line intensities in both observations are close 
to the noise level and so we leave these as undetected.

We also find two U lines in this overlapped spectral region 
for which ZM93 did not give any identification. 
In contrast ZM93 find 4 U lines for  which we did not provide an identification.
However, the antenna temperatures of the six U lines are $\sim 0.1$ K, 
which is close to noise level.
Follow-up observations would be necessary for 
confirmation of the identification of the weak lines.  
  
The identified 50 emission lines are found to originate from a 
total of 16 species and 
their isotopic variants.
The strongest molecular transition between 138.3 to 150.7 GHz is
the J=$3-2$ transition of the CS molecule, followed by
the lines of
the $\rm H_2CO$, $\rm CH_3OH$, $\rm CH_3CN$, and $\rm SO_2$.

An interesting point is that a considerable number of the detected lines 
are from large organic molecules  
such as $\rm CH_3OH$ (30), $\rm ^{13}CH_3OH$ (4), $\rm CH_3OCH_3$ (20),
$\rm HCOOCH_3$ (16), $\rm C_2H_5CN$ (15), and $\rm CH_3CN$ (7)
(the numeric in parenthesis is the number of the detected transitions of the
molecule).
About 80 \% of the identified lines are from the transitions of these
organic molecules.

\subsection{The LTE Rotation Diagram Analysis and Individual Molecules}
We perform the standard LTE rotational diagram analysis to derive 
rotation temperatures ($\rm T_{rot}$) and column densities ($\rm N$) 
of several molecules for which more than four transitions are detected.
The derived physical quantities are compared with results 
by other studies in different wavelength bands toward Orion-KL.

The equation for the LTE rotation 
diagram analysis (e.g., Turner 1991) is given as
$$
log L = log {{3k \int T_A^* dv}\over{8 \eta_B \pi^3 \nu S\mu_i^2 g_I g_k}} 
= log{N\over Q_{rot}} -{{E_u~log~e} \over {k~T_{rot}}}, \eqno (1)
$$
where k is the Boltzmann constant, $\int T_A^* dv$ the integrated intensity,
$\eta_B$ the beam efficiency of the telescope, 
$\nu$ the rest frequency of the spectrum, S the line strength, $\mu_i$ 
the relevant dipole moment, $g_I$ the reduced nuclear spin weight,
$g_k$ the K-level degeneracy, $E_u$ the upper state energy of the transition,
$Q_{rot}$ the rotational partition function, and N the column density.  

The values for  $g_I$, $g_k$, $E_u/k$, $S\mu_i^2$, $\int T_A^* dv$, and $log L$
for each molecule and their references are given in Table $3 - 10$. 
The partition functions $Q_{rot}$ are 
${1 \over 2} [{{\pi (kT_{rot})^3} \over {h^3ABC}}]^{1/2}$ for $\rm H_2CO$ and 
$\rm SO_2$,
${1 \over 3} [{{\pi (kT_{rot})^3} \over {h^3AB^2}}]^{1/2}$ for $\rm CH_3CN$,
$ [{{\pi (kT_{rot})^3} \over {h^3ABC}}]^{1/2}$ for $\rm CH_3OCH_3$ and 
$\rm C_2H_5CN$,
and $2 [{{\pi (kT_{rot})^3} \over {h^3ABC}}]^{1/2}$ for $\rm CH_3OH$ and 
$\rm HCOOCH_3$, where
A, B, and  C are the rotational constants (Blake et al. 1986; Turner 1991).

The above equation is derived under the assumption
that all lines
are optically thin, all level populations are characterized by a single
$T_{rot}$ (LTE), $\rm T_{rot} >> T_{bg}$ (background temperature), 
and that the Rayleigh-Jeans approximation is valid for all transitions 
(see Turner 1991). 
If these assumptions
are valid and the line 
identifications are correct, the distribution of $E_u/k$ versus log L 
of the data would 
exhibit a linear correlation, and the rotational temperature and the column density
can be determined from the 
slope [$-(log e)/T_{rot}$] and intercept [$log (N/Q_{rot})$ at $E_u = 0$]
through a linear least squares fit of the data. 

On the other hand, if the distribution of $E_u/k$ 
versus log L of the data in the rotation diagram deviates greatly from 
the linear correlation,
the assumptions of low optical depth and LTE may not be valid 
and requires further consideration in deriving $\rm T_{ex}$ and N from the data.
Moreover an incorrect line identification may also result in a
large  deviation of the data point from the correlation.
Therefore, the LTE rotational diagram analysis itself  
can be useful for checking secure identification of spectral lines.
 
We applied this analysis for 7 molecules and one isotopic variant: 
$\rm CH_3OH$ ($\rm ^{13}CH_3OH$), $\rm HCOOCH_3$, $\rm CH_3OCH_3$, $\rm C_2H_5CN$,
$\rm SO_2$, $\rm CH_3CN$, and $\rm H_2CO$.
The transitions of 4 molecules [$\rm CH_3OH$, $\rm HCOOCH_3$, 
$\rm SO_2$, and $\rm CH_3CN$] and one isotopic variant ($\rm ^{13}CH_3OH$) 
are found to exhibit a
linear distribution in the rotation diagram, allowing us to introduce the estimates of 
the physical quantities of the molecules and 
compare our results with those from other surveys.
On the other hand,  the rotational diagrams for three molecules:
$\rm CH_3OCH_3$, $\rm C_2H_5CN$ and $\rm H_2CO$ show 
scattered non-linear distributions.
Therefore, the rotation temperatures and the column densities for these 
molecules are not derived.

Fig. 3 shows the diagrams for 6 molecules and one isotope variant 
used in the analysis. The diagram for $\rm H_2CO$  is not shown 
because of the large scatter seen in the data.    
In the following paragraphs we discuss the results of the rotational 
diagram analysis in sequence of increasing number of detected transitions
of the molecule.

{\bf CH$_3$OH--} Methanol is one of most widely observed species 
in molecular clouds and star-forming regions. 
The number of transitions detected (30) is
the largest in our observations (Table 3).
We find 19 new detections from a total of 30 methanol lines identified
in our survey.
The data distribution in the rotation diagram shows a fairly linear distribution
with some scatter. In the rotation diagram analysis we use 26 transitions 
for which $\rm E_u/k$ and $\rm S\mu^2$ are known.  
We removed a single data point from the [$3(1)-2(1)$ A-] transition 
in the analysis which shows a large deviation in the diagram (Fig. 3).
The linear least squares fit to the data indicates 
a fairly good correlation (correlation coefficient; $r=-0.86$), 
yielding  results of $\rm T_{ex}=160_{-17}^{+21}$ K
and $\rm N=2.1_{-0.7}^{+1.1}\times 10^{16}~cm^{-2}$ which  
are close to 3 mm results ($\rm T_{ex}=192$ K
and $\rm N=1.6\times 10^{16}~cm^{-2}$) obtained by Turner (1991). 

Four transitions of the rare isotopic species, $\rm ^{13}CH_3OH$, 
are also detected with three new detections 
(Table 4). The linear least squares fit
to the data, excluding a single deviated point [3(-1)-2(-1) E] results in 
a nearly perfect correlation ($r=-0.998$) 
(Fig. 3), producing $\rm T_{ex}=40_{-2}^{+3}$ K
and $\rm N=1.2_{-0.2}^{+0.2}\times 10^{14}~cm^{-2}$. 
Using the terrestrial value of $\rm ^{12}C/^{13}C=89$ may provide an
alternate (and more reliable) estimate ($\rm N=1.1\times 10^{16}~cm^{-2}$)
of the column density of Methanol, which 
is close to the value determined using the LTE rotational method 
within the uncertainty.

{\bf CH$_3$OCH$_3$--} Dimethyl ether is one of a few 
interstellar molecules with two internal rotors and
we detect twenty transitions from this molecule in our survey (Table 5).
In contrast to Methanol, there are no newly identified lines.
The rotation diagram shows very scattered distribution of the data,
indicating no  linear correlation between $\rm E_u/k$ and log L (Fig. 3).  
This prevents us from inferring confident values of $\rm T_{rot}$ and N.
Note that Johansson et al. (1984),
Blake et al (1986), and Turner (1991) have deduced 
$\rm T_{ex}=63 - 91$ K and $\rm N=1.3 - 3.0 \times 10^{15}~cm^{-2}$ 
from their 3 mm data.  

{\bf HCOOCH$_3$--} Sixteen transitions of Methyl formate are detected (Table 6),
with 8 lines believed to be `first' detections. 
Turner (1991) has shown that this species has two components in the rotation 
diagram and
our data in the rotational diagram also seem to exhibit two 
components (Fig. 3).
The rotation analysis for one component 
yields $\rm T_{ex}=21.8_{-4.4}^{+7.2}$ K and
$\rm N=2.3_{-1.5}^{+4.9} \times 10^{15}~cm^{-2}$.
This rotational temperature is very close to that found by Turner 
(1991), but the column density is one order of magnitude higher than his result. 
The other, cooler component, shows $\rm T_{ex}=18.1_{-3.6}^{+6.0}$ K 
and $\rm N=2.6_{-1.7}^{+5.5} \times 10^{14}~cm^{-2}$  
has not been seen by any other studies.

{\bf C$_2$H$_5$CN--} Fifteen transitions of Ethyl cyanide are detected (Table 7). 
Twelve lines are considered to be new detections.
Interestingly, seven transitions seem to be K-type doubler lines whose 
frequencies are the same.  
The distribution of the data is not a simple linear appearance.
So, the rotational diagram method does not seem to be applicable
to derive $\rm T_{ex}$ and N of this molecule.

{\bf SO$_2$--} Sulfur dioxide is detected in seven transitions 
(Table 8) of which four are newly detected lines. 
Five strong lines are found to consist of two clear velocity components,  
a relatively narrow and bright component of 
$\rm V_{LSR}\approx 6 - 8~km~s^{-1}$ and $\rm \Delta V_{FWHM}\approx 
14 - 31~km~s^{-1}$, and  
a wide and weak component of $\rm V_{LSR}\approx 10 - 18~km~s^{-1}$ and 
$\rm \Delta V_{FWHM}\approx 15 - 34~km~s^{-1}$.
The data  of  Log L and $\rm E_u/k$ for the former component
show a good correlation ($r=-0.92$) 
with a slight scatter in the rotation diagram (Fig. 3). 
The analysis yields $\rm T_{ex}=120_{-20}^{+21}$ K and 
$\rm N=1.7_{-0.1}^{+1.5} \times 10^{16}~cm^{-2}$.   
Considering the uncertainty, our $\rm T_{ex}$ is fairly comparable
to the value (138.5 K) found in the 3 mm observations
by Turner (1991) and the value ($\sim 140$ K) in 2 mm observations
obtained by ZM93. The column density is also in close agreement with 
the results ($\rm N\approx 1.5\times10^{16}~cm^{-2}$) by Turner (1991) and
($\rm N\approx 2.0\times10^{16}~cm^{-2}$) by ZM93.  
The rotation diagram for the wide and weak component which we did not include
in the Fig. 3 produces a very scattered distribution 
and we do not to derive $\rm T_{ex}$ and the N.

{\bf CH$_3$CN--} Methyl cyanide emissions  are detected in the seven 
K-components
from the J=$8-7$ transition over very narrow frequency range of about 700 MHz (Table 9). 
One transition [8(6)-7(6)] is a new detection.
Previously different investigators have suggested several different values 
for the  $\rm T_{ex}$ and N primarily due to the large scatter in the data. 
The overall distribution of our data points in the 
rotation diagram shows some scatter in the correlation ($r=-0.61$) (Fig. 3).  
The least squares fit results in  $\rm T_{ex}=273_{-101}^{+381}$ K and 
$\rm N=1.3_{-0.1}^{+5.1} \times 10^{15}~cm^{-2}$.
The $\rm T_{ex}$ is virtually the same as the value (274 K) for the hot core
component by Blake et al. (1986) and the column density for the component
is close to their estimate.

{\bf H$_2$CO--} Formaldehyde has 
three detected transitions (Table 10). 
All of the transitions have been previously observed by other studies.
The line shapes of all spectra
clearly exhibit two components, one a narrow bright emission component   of
$\rm V_{LSR}\approx 8.6 - 8.9~km~s^{-1}$ and 
$\rm \Delta V_{FWHM}\approx 5 - 6 ~km~s^{-1}$,
and the other a broad faint emission component (presumably plateau emission) of 
$\rm V_{LSR}\approx 7.0 - 8.3~km~s^{-1}$ and
$\rm \Delta V_{FWHM}\approx 15 - 20 ~km~s^{-1}$.
Due to the high degree of scatter in the correlation we have not included
the rotation diagram in Fig. 3 and also do not derive a rotational temperature
and total column density for the two components.
 
In summary, the LTE rotational diagram analysis is fairly useful to derive 
important physical quantities ($\rm T_{ex}$ and N) from four molecules and one isotopic variant, 
each with more than 4 detected transitions. 
All the values are comparable to the estimates in 
different frequency regions by other researchers.
The estimated values are compared  with those from 
other studies in Table 11.
 
\subsection{Other Interesting Lines}

{\bf CS Lines--} Of the observed emission lines the CS molecule  stands out
in that the
CS (3$-$2) emission is the strongest among the detected transitions and
its three rare isotopes such as $\rm ^{13}CS$,
$\rm  C^{34}S$, and $\rm  C^{33}S $ are also strongly detected in our survey
(Table 12).
The main isotope CS ($3-2$) line appears to contain several components,
among which a bright ridge component of $\rm V_{LSR}\approx 5.5~km~s^{-1}$ and
$\rm \Delta V_{FWHM}\approx 8.2~km~s^{-1}$, and a broad plateau component of
$\rm V_{LSR}\approx 6.5~km~s^{-1}$ and 
$\rm \Delta V_{FWHM}\approx 21.5~km~s^{-1}$ are included.
The three rare-isotope species seem to trace a
ridge component of $\rm V_{LSR}\approx 8\sim 9~km~s^{-1}$ and
$\rm \Delta V_{FWHM}\approx 5~km~s^{-1}$.
Comparison of the CS emission with emissions from the isotope
variants can give the CS optical depth.
The ratios of integrated intensities between $\rm  C^{32}S$ and $\rm  C^{33}S$,
and between $\rm  C^{32}S$ and $\rm  C^{34}S$ are about
36.6 and 5.5, respectively. These are significantly different from
the terrestrial values of $\rm ^{32}S/^{33}S=80$ and $\rm ^{32}S/^{34}S=18$,
meaning that the CS (3$-$2) emission is optically thick.
This is also inferred from the fact that
the ratio (about 11.9) of the integrated intensities
between CS and $\rm ^{13}CS$  is small compared with the
terrestrial value of $\rm ^{12}C/^{13}C=89$.
Assuming $\rm T_{ex}=65$ K and $\rm ^{32}S/^{34}S=18$ which are adopted
from Sutton et al. (1995),
we obtain $\rm N(C^{34}S)\approx 1.1\times 10^{13}~cm^{-2}$ and
$\rm N(CS)\approx 2.0\times 10^{14} ~cm^{-2}$ using the LTE method
(e.g., Lee, Minh, \& Irvine 1993).
These estimates are one order of magnitude smaller than those of Sutton et al. (1995).
This difference may be the result of the larger beam dilution in our survey.

{\bf Recombination Lines--} The recombination lines 
in our survey frequency region 
are $\rm H35\alpha$ (147046.9 MHz), $\rm H44\beta$ (144474.0 MHz),
$\rm H50\gamma$ (144678.9 MHz), $\rm He35\alpha$ (147106.8 MHz),
$\rm He44\beta$ (144532.9 MHz), and $\rm He50\gamma$ (144737.9 MHz).
The most plausible candidate of the radio recombination lines
may be the U147112.9 line which has close frequency
to $\rm He35\alpha$, but the frequency difference between two
is somewhat large ($\sim 6$ MHz). So we leave the U147112.9 line unidentified.
There is no clear evidence of detection for other recombination lines
in our survey.

{\bf Unidentified (U) Lines--}
A total of 32 U lines toward
Orion-KL are found in our survey. Several lines (e.g., U143006.7 and U143263.3)
are tentative detections because of low S/N ratios
of the spectra. 
We note that U139561.9, U146372.4, U146622.4, U146984.5, U147759.8, and U147943.7
are quite bright ($\rm T_A^* \ga 0.3~K$).
The frequency, $\rm \Delta V_{FWHM}$, antenna temperature, and integrated intensity
of the U lines are given in Table 13 and marked with `U' on the spectra in Fig. 2.
All U lines have 
$\rm \Delta V_{FWHM}\approx 2 - 18~km~s^{-1}$ and seem to originate from
various components such as ridge, hot core, and plateau.
The U147943.7 line has two velocity components,
presumably a hot core component ($\rm \Delta V_{FWHM}\approx 5.1~km~s^{-1}$) and 
a plateau component ($\rm \Delta V_{FWHM}\approx 13.4~km~s^{-1}$).
The identities of numerous  U lines  may be either from unknown  transitions of
the known molecules with complex rotational structure
or from any transitions of unknown molecules.
We leave the identifications of these lines to be explored in a future study.

\subsection{Regional Origin of the Molecular Species}
The spectral lines of each molecular species can often trace physically different 
regions of molecular clouds,  and so are useful for establishing the origin
of the species and understanding the astro-chemical processes inside 
molecular clouds. 
Recently, Sutton et al (1995) have surveyed 5 distinct regions of the Orion molecular
clouds with high spatial resolution (about $14''$ beam ) and
characterized the kinematic properties and main tracers of the 
four distinguished spatial and kinematical components such as extended ridge, hot core,
compact ridge, and plateau.
Their data suggest that the extended ridge shows $\rm V_{LSR}
\approx 9.8~km~s^{-1}$ and $\rm \Delta V_{FWHM} \approx 2.5~km~s^{-1}$,
and the hot core $\rm V_{LSR}
\approx 7.0~km~s^{-1}$ and $\rm \Delta V_{FWHM} \approx 8.0~km~s^{-1}$,
the compact ridge $\rm V_{LSR}
\approx 6.6~km~s^{-1}$ and $\rm \Delta V_{FWHM} \approx 8.0~km~s^{-1}$,
and the plateau $\rm V_{LSR}
\approx 5.5\sim 9.8~km~s^{-1}$ and $\rm \Delta V_{FWHM} \approx 
10 \sim 20~km~s^{-1}$.
The beam of the TRAO telescope at 140 GHz is more than 3 times bigger 
than the JCMT beam at 330 GHz, and so  encompasses nearly 
all components by Sutton et al. (1995). 
By comparing their kinematical values of these distinct components 
in the Orion clouds with our data, 
we  are able to determine which component a given molecular species
traces. For this purpose, we obtained  
the $\rm V_{LSR}$ and $\rm \Delta V_{FWHM}$ of each detected transition
from all species
from Gaussian fits to the spectral lines (Table 11), which are also plotted in Fig. 4. 
The values for the molecules with 
multiple transitions are given in average. 
According to the kinematic values of the species, 
$\rm CH_2CHCN$
is thought to be mainly from the extended ridge and
$\rm C_2H_5CN$, $\rm CH_3CN$, $\rm H_2CO$,    
$\rm SO_2$, $\rm NO$, $\rm CS$, $\rm HC_3N$, and $\rm OCS$
from the plateau.
The kinematic values of the remaining species are similar 
to the values of the hot core or the compact ridge, and so 
it is difficult to discriminate where they are from, between the two regions.

\section{Summary}
We present the results from a spectral line survey toward Orion-KL 
in the frequency range from 138.3 to 150.7 GHz carried out using 
the 14 m radio telescope of TRAO.

A total of 149 molecular line spectra are detected from the survey.
Line identification for the detected lines is performed using 
Lovas catalog (Lovas 1992) which is the collection of 
the observed molecular lines, and  
JPL (Pickett et al. 1998), Kawaguchi (1999), Anderson et al. (1992), 
Xu \& Lovas (1997) and Tsunekawa et al. (2000)  catalogs which are the collections of
laboratory measurements or theoretical calculations of molecular transitions.

We identify fifty lines 
which  have been previously reported by other studies, and 99 lines  
which are found to be new detections in the interstellar medium. 
Sixty-seven among the new lines are from molecular transitions
with known identifications, however, 
32 spectral lines  remain ``$\it Unidentified$''.
No clear evidence for detection of the recombination lines of H and He 
is found in our survey.   

The detected emissions in the 2mm band arise
from a total of 16 molecular species 
and their isotopic variants. 
It is important to note that about 80 \% of the detected lines are from transitions
of the large organic molecules such as $\rm CH_3OH$, $\rm CH_3OCH_3$, 
$\rm HCOOCH_3$, $\rm C_3H_5CN$, and $\rm CH_3CN$. 
We found that the LTE rotation diagram analysis is useful as a check on the
line identification and is used for deducing physical quantities of 
four molecular species and an isotopic variant, namely 
[$\rm CH_3OH$ ($\rm ^{13}CH_3OH$), $\rm HCOOCH_3$, 
$\rm CH_3CN$, and $\rm SO_2$]. 
Using this method we derive their rotation temperatures ($\rm 10\sim 270~K$)
and column densities ($\rm 0.2\sim 20\times 10^{15}~cm^{-2}$). 
These values are fairly comparable to the estimates 
by other studies in other frequency regions, indicating our line 
identifications are fairly secure.
Relying on the kinematic values of the molecular species, 
it is found that   
emission of $\rm CH_2CHCN$ arises from the extended ridge and
emissions of $\rm C_2H_5CN$, $\rm CH_3CN$, $\rm H_2CO$,    
$\rm SO_2$, $\rm NO$, $\rm CS$, $\rm HC_3N$, and $\rm OCS$
originate in the plateau.

\acknowledgments
We would like to express sincere thanks to Kentarou Kawaguchi 
for providing his unpublished spectral line catalog   
and suggesting useful comments.
The discussion with Masatoshi Ohishi aided 
in the line identification is also greatly acknowledged.
We are in debt to Edwin Bergin for his careful reading of our
manuscript.
We would also like to thank an anonymous referee for his (or her) 
useful comments.
This research is  supported  by Star Project Grant No. 97-NQ-01-01-A and 
No. 98-NQ-01-01-A of the Korea Astronomy Observatory, which is sponsored
by the Ministry of Science and Technology. 

\vfill\eject

\clearpage

\clearpage


\begin{figure}
\noindent{\bf Fig. 1. ---} All spectra obtained from 138.3 to 150.7 GHz 
toward Orion-KL. Strong lines are designated their identifications.
`U' is to mark `Unidentified' line. 
\end{figure}

\begin{figure}
\noindent{\bf Fig. 2. --- } 
Line identifications of the spectra detected from 138.3 to 150.7 GHz
toward Orion-KL. The spectral resolution is 1 MHz.
The rest frequency scale for all spectra 
is determined by  assuming $\rm V_{LSR}=8.4~km~s^{-1}$. 

\end{figure}

\begin{figure}
\noindent{\bf Fig. 3. --- } Rotation Diagrams for the molecules
with at least four detected transitions.
The error bar is 1 $\sigma$. The data marked with 
open circles in the diagrams for $\rm CH_3OH$ and $\rm ^{13}CH_3OH$ 
are excluded in the linear least squares fit of the data.  
The diagrams for $\rm CH_3OCH_3$ and $\rm C_2H_5CN$ show little 
correlation between $\rm E_u/k$ and log L of the data, and so 
the linear least squares fit is not applied to the molecules.

\end{figure}

\begin{figure}
\noindent{\bf Fig. 4. --- } Plot of $\rm V_{LSR}$ versus FWHM 
of the molecules  
detected from from 138.3 to 150.7 GHz toward Orion-KL. 
The values for the molecules with multiple transitions are given in average. 

\end{figure}
\end{document}


\pagestyle{empty}

\begin{deluxetable}{rllll}
\tablewidth{40pc}
\tablecaption{Previous spectral line surveys toward the Orion-KL region}
\tablehead{
        \colhead{observed Frequency}  &\colhead{Telescope} &
        \colhead{Reference}           &\colhead{Comments} 
        }
\startdata
 72~-~~91 GHz  & Onsala 20 m  &  Johansson et al. 1984           &                        \nl
~70 - 115 GHz  & NRAO 12 m    &  Turner 1989                     &                        \nl
150 - 160 GHz  & FCRAO 14 m   &  Ziurys \& McGonagle 1993        &                        \nl
215 - 247 GHz  & OVRO 10.4 m  &  Sutton et al. 1985              &                        \nl
216 - 242 GHz  & OVRO Array   &  Blake et al.  1996              &                        \nl
247 - 263 GHz  & OVRO 10.4 m  &  Blake et al. 1986               &                        \nl
257 - 273 GHz  & JCMT 15 m    &  Greaves \& White 1991           &                        \nl 
330 - 360 GHz  & NRAO 12 m    &  Jewell et al. 1989              &                        \nl 
325 - 360 GHz  & CSO 10.4 m   &  Schilke et al. 1997             &                        \nl 
334 - 343 GHz  & JCMT 15 m    &  Sutton et al. 1995              &                        \nl 
607 - 725 GHz  & CSO 10.4 m   &  Schilke et al. 1999             &                        \nl 
780 - 900 GHz  & CSO 10.4 m   &  Schilke et al. 1999             & in progress            \nl 
190 - 900 GHz  & CSO 10.4 m   &  Serabyn \& Weisstein 1995       & Low resolution (200 MHz)\nl 
\enddata			      


\end{deluxetable}
\clearpage
 
\begin{deluxetable}{cllrll}
\tablecaption{Identifications of Spectral Lines Detected Toward Orion-KL}
\tablehead{
        \colhead{Frequency}      &\colhead{Species} & \colhead{Transition}     &
        \colhead{$\rm T_A^*$\tablenotemark{a}}&\colhead{Reference\tablenotemark{b}}   &
        \colhead{Comments\tablenotemark{c}}  \nl 
        \colhead{(MHz)}      &\colhead{} & \colhead{}     &
        \colhead{(K)}&\colhead{}   &
        \colhead{}   
        }
\startdata
138259.9& $\rm U       $    &                                            &  0.10&          &         \nl
138343.2& $\rm U $          &                                            &  0.09&          &         \nl
138351.1& $\rm  C_2H_5CN$   &16(1,16)-15(1,15)                           &  0.29& JPL      & NDT     \nl
138396.4& $\rm U $          &                                            &  0.13&          &         \nl
138739.3& $\rm ^{13}CS$     & J=3-2                                      &  0.56& Lovas    &         \nl
139256.7& $\rm U $          &                                            &  0.09&          &         \nl 
139335.9& $\rm  C_2H_5CN$   &16(0,16)-15(0,15)                           &  0.17& JPL      & NDT     \nl
139355.1& $\rm SO_2    $    & 5(3,3)-6(2,4)                              &  0.26& JPL      & NDT     \nl
139416.9& $\rm U     $      &                                            &  0.10&          &         \nl
139432.5& $\rm U         $  &                                            &  0.14&          &         \nl
139436.0& $\rm U        $   &                                            &  0.14&          &         \nl
139474.5& $\rm SO_2$        & 26(7,19)-27(6,22)                          &  0.12& JPL      & NDT     \nl
139483.5& $\rm H_2CS$       & 4(1,3)-3(1,2)                              &  0.41& Lovas    &         \nl
139500.4& $\rm CH_3OCH_3$   & 9(3,6)-9(2,7) AE                           &  0.10& JPL      & NDT     \nl
139503.7& $\rm CH_3OCH_3$   & 9(3,6)-9(2,7) EE                           &  0.15& JPL      & NDT     \nl
139506.9& $\rm CH_3OCH_3$   & 9(3,6)-9(2,7) AA                           &  0.16& JPL      & NDT     \nl
139561.9& $\rm U $          &                                            &  0.32&          &         \nl
139582.1& $\rm U $          &                                            &  0.13&          &         \nl
139862.3& $\rm U $          &                                            &  0.08&          &         \nl
139878.4& $\rm U        $   &                                            &  0.10&          &         \nl
139903.0& $\rm CH_3OH   $   & unassigned                                 &  0.09& TUTT     & NDT     \nl
140033.1& $\rm CH_3OH   $   & 23(-2)-23(1) E                             &  0.17& Lovas    &         \nl
140127.5& $\rm H_2CCO$      & 7(1,7)-6(1,6)                              &  0.10& Lovas    &         \nl
140151.1& $\rm CH_3OH   $   & $v_t=0$, 18(0)-18(-1) E                    &  0.29& Lovas    &         \nl
140166.6& $\rm CH_3OH   $   & unassigned                                 &  0.10& TUTT     & NDT     \nl
140174.6& $\rm U        $   &                                            &  0.08&          &         \nl
140306.2& $\rm SO_2     $   & 6(2,4)-6(1,5)                              &  1.31& Lovas    &         \nl
140839.5& $\rm H_2CO    $   & 2(1,2)-1(1,1)                              &  3.14& Lovas    &         \nl
141260.4& $\rm HCOOCH_3 $   & 11(3,8)-10(3,7) A                          &  0.41& Lovas    &         \nl
141402.5& $\rm H_2CCO  $    & 7(3,4)-6(3,3)\& 7(3,5)-6(3,4)              &  0.11& JPL      & NDT     \nl
141441.3& $\rm CH_3OH    $  & $v_t=0$, 0(0)-1(1) E                       &  0.28& XL       & NDT     \nl
141597.1& $\rm ^{13}CH_3OH$ & $v_t=0$, 3(-1)-2(-1) E                     &  0.29& XL       & NDT     \nl
141602.5& $\rm ^{13}CH_3OH$ & $v_t=0$, 3(0)-2(0) A                       &  0.26& Lovas    &         \nl
141623.5& $\rm ^{13}CH_3OH$ & $v_t=0$, 3(2)-2(2) E \& 3(-2)-2(-2) E      &  0.32& XL       & NDT     \nl
141629.2& $\rm ^{13}CH_3OH$ & $v_t=0$, 3(1)-2(1) E                       &  0.28& XL       & NDT     \nl
141646.5& $\rm U   $        &                                            &  0.13&          &         \nl
141653.0& $\rm  HCOOCH_3  $ & 11(2,9)-10(2,8) E                          &  0.56& JPL      & NDT     \nl 
141667.0& $\rm  HCOOCH_3 $  & 11(2,9)-10(2,8) A                          &  0.38& JPL      & NDT     \nl
141829.1& $\rm  CH_3OCH_3$  & 8(3,5)-8(2,6) EA                           &  0.17& JPL      & NDT     \nl
141832.3& $\rm  CH_3OCH_3$  & 8(3,5)-8(2,6) EE                           &  0.37& JPL      & NDT     \nl
141835.5& $\rm  CH_3OCH_3$  & 8(3,5)-8(2,6) AA                           &  0.18& JPL      & NDT     \nl
141983.7& $\rm  U         $ &                                            &  0.36&          &         \nl
142054.4& $\rm  CH_3OH   $  & unassigned                                 &  0.14& TUTT     & NDT     \nl
142346.3& $\rm  C_2H_5CN$   &16(2,15)-15(2,14)                           &  0.21& JPL      & NDT     \nl
142735.1& $\rm HCOOCH_3   $ & 13(1,13)-12(1,12) A                        &  0.33& Lovas    &         \nl
142815.5& $\rm HCOOCH_3  $  & 13(0,13)-12(0,12) E                        &  0.20& JPL      & NDT     \nl
142924.5& $\rm HCOOCH_3  $  & 13(1,13)-12(0,12) E                        &  0.23& JPL      & NDT     \nl
143006.7& $\rm  U        $  &                                            &  0.20&          &         \nl
143057.1& $\rm  SO_2$       & 16(2,14)-16(1,15)                          &  0.97& Lovas    &         \nl
143108.3& $\rm  CH_3OH$     & $v_t=0$, 17(0)-17(-1) E                    &  0.18& XL       & NDT     \nl
143169.5& $\rm  CH_3OH$     & $v_t=0$, 7(3)-8(2) E                       &  0.97& XL       & NDT     \nl
143240.5& $\rm  HCOOCH_3  $ &  12(1,11)-11(1,10) A                       &  0.34& JPL      & NDT     \nl
143263.3& $\rm  U         $ &                                            &  0.07&          &         \nl
143337.7& $\rm  C_2H_5CN$   &16(7,9)-15(7,8) \& 16(7,10)-15(7,9)         &  0.33& JPL      & NDT     \nl
143343.9& $\rm  C_2H_5CN$   &16(9,8)-15(9,7) \& 16(9,7)-15(9,6)          &  0.19& JPL      & NDT     \nl
143357.2& $\rm  C_2H_5CN$   &16(6,10)-15(6,9) \& 16(6,11)-15(6,10)       &  0.28& JPL      & NDT     \nl
143360.4& $\rm  C_2H_5CN$   &16(10,7)-15(10,6) \& 16(10,6)-15(10,5)      &  0.25& JPL      & NDT     \nl
143383.0& $\rm  C_2H_5CN$   &16(11,6)-15(11,5) \& 16(11,5)-15(11,4)      &  0.09& JPL      & NDT     \nl
143407.2& $\rm  C_2H_5CN$   &16(5,12)-15(5,11) \& 16(5,11)-15(5,10)      &  0.32& JPL      & NDT     \nl
143410.8& $\rm  C_2H_5CN$   &16(12,5)-15(12,4) \& 16(12,4)-15(12,3)      &  0.22& JPL      & NDT     \nl
143507.0& $\rm  C_2H_5CN$   &16(4,13)-15(4,12)                           &  0.22& Lovas    &         \nl
143529.2& $\rm  C_2H_5CN$   & 16(3,14)-15(3,13)                          &  0.26& Lovas    &         \nl
143535.3& $\rm  C_2H_5CN$   & 16(4,12)-15(4,11)                          &  0.22& Lovas    &         \nl
143599.4& $\rm  CH_3OCH_3$  &  7(3,4)-7(2,5) AE                          &  0.15& Lovas    &         \nl
143603.0& $\rm  CH_3OCH_3$  &  7(3,4)-7(2,5) EE                          &  0.18& Lovas    &         \nl
143606.2& $\rm  CH_3OCH_3$  &  7(3,4)-7(2,5) AA                          &  0.17& Lovas    &         \nl
143618.1& $\rm  C_2H_5OH $  &  28(6,23)-29(2,27) gauche                  &  0.14& JPL      & NDT     \nl
143821.6& $\rm  U        $  &                                            &  0.20&          &         \nl
143865.8& $\rm  CH_3OH  $   & $v_t=0$, 3(1)-2(1) A                       &  0.66& Lovas    &         \nl
144267.0& $\rm  U       $   &                                            &  0.11&          &         \nl
144351.4& $\rm  U        $  &                                            &  0.14&          &         \nl
144370.2& $\rm  U        $  &                                            &  0.13&          &         \nl
144572.0& $\rm  CH_3OH  $   & $v_t=2$, 3(0)-2(0) A                       &  0.09& ALH      & NDT     \nl
144583.9& $\rm  CH_3OH  $   & $v_t=2$, 3(-1)-2(-1) E                     &  0.09& ALH      & NDT     \nl
144589.8& $\rm  CH_3OH   $  & $v_t=1$, 3(1)-2(1) A                       &  0.14& XL       & NDT     \nl
144617.1& $\rm  C^{34}S $   &  J=3-2                                     &  1.31& Lovas    &         \nl
144728.7& $\rm  CH_3OH  $   & $v_t=1$, 3(-2)-2(-2) E \& 3(2)-2(2) A      &  0.31& XL       & NDT     \nl
144734.5& $\rm  CH_3OH  $   & $v_t=1$, 3(1)-2(1) E                       &  0.41& XL       & NDT     \nl
144750.2& $\rm  CH_3OH  $   & $v_t=1$, 3(-1)-2(-1) E                     &  0.23& XL       & NDT     \nl
144762.2& $\rm  CH_2CHCN-v2$& 15(2,13)-14(2,12)                          &  0.21& Kawaguchi& NDT     \nl
144768.2& $\rm  CH_3OH  $   & $v_t=1$, 3(0)-2(1) A                       &  0.18& XL       & NDT     \nl
144828.0& $\rm  DCN      $  &  J=2-1   F'=2-1                            &  0.96& Lovas    &         \nl
144859.0& $\rm  CH_3OCH_3$  &  6(3,3)-6(2,4) EE                          &  0.17& JPL      & NDT     \nl
144878.6& $\rm  CH_3OH  $   & $v_t=1$, 3(1)-2(1) A                       &  0.19& XL       & NDT     \nl
145093.6& $\rm  CH_3OH   $  & $v_t=0$, 3(0)-2(0) E                       &  1.82& Lovas    &         \nl
145097.5& $\rm  CH_3OH   $  & $v_t=0$, 3(-1)-2(-1) E                     &  2.06& Lovas    &         \nl
145103.2& $\rm  CH_3OH   $  & $v_t=0$, 3(0)-2(0) A                       &  1.82& Lovas    &         \nl
145124.4& $\rm  CH_3OH   $  & $v_t=0$, 3(2)-2(2) A                       &  1.50& Lovas    &         \nl
145126.4& $\rm  CH_3OH   $  & $v_t=0$, 3(2)-2(2) E \&    3(-2)-3(-2) E   &  2.00& Lovas    &         \nl
145131.9& $\rm  CH_3OH   $  & $v_t=0$, 3(1)-2(1) E                       &  1.70& Lovas    &         \nl
145133.5& $\rm  CH_3OH   $  & $v_t=0$, 3(2)-2(2) A                       &  1.74& Lovas    &         \nl
145560.9& $\rm  HC_3N    $  &  J=16-15                                   &  1.01& Lovas    &         \nl
145602.9& $\rm  H_2CO    $  &  2(0,2)-1(0,1)                             &  2.45& Lovas    &         \nl
145680.4& $\rm CH_3OCH_3$   &  5(3,2)-5(2,3) EE                          &  0.14& Lovas    &         \nl
145755.6& $\rm  C^{33}S$    &  J=3-2 F=9/2-7/2 \& F=7/2-5/2              &  0.19& Lovas    &         \nl
145766.1& $\rm  CH_3OH   $  & $v_t=0$, 16(0)-16(-1) E                    &  0.29& XL       & NDT     \nl
145876.2& $\rm  U       $   &                                            &  0.15&          &         \nl
145946.8& $\rm  OCS     $   &  J=12-11                                   &  0.80& Lovas    &         \nl
146120.0& $\rm  C_2H_5CN$   &16(2,14)-15(2,13)                           &  0.28& JPL      & NDT     \nl
146129.6& $\rm  U       $   &                                            &  0.17&          &         \nl
146368.3& $\rm  CH_3OH  $   & $v_t=0$, 3(1)-2(1) A                       &  0.96& XL       & NDT     \nl
146372.4& $\rm  U  $        &                                            &  0.42&          &         \nl
146550.1& $\rm  SO_2$       &  10(4,6)-11(3,9)                           &  0.32& JPL      & NDT     \nl
146605.5& $\rm  SO_2     $  &  4(2,2)-4(1,3)                             &  0.99& JPL      & NDT     \nl
146618.8& $\rm  CH_3OH   $  & $v_t=0$, 9(0)-8(1) A                       &  1.23& XL       & NDT     \nl
146622.4& $\rm  U        $  &                                            &  0.74&          &         \nl
146894.5& $\rm  C_2H_5CN$   &17(1,17)-16(1,16)                           &  0.28& JPL      & NDT     \nl
146969.0& $\rm  CS       $  &  J=3-2                                     &  6.17& Lovas    &         \nl
146977.7& $\rm HCOOCH_3$    &  12(3,10)-11(3,9) E                        &  0.63& Lovas    &         \nl
146984.5& $\rm  U        $  &                                            &  0.38&          &         \nl
146988.0& $\rm HCOOCH_3$    &  12(3,10)-11(3,9) A                        &  0.44& Lovas    &         \nl
147024.9& $\rm  CH_3OCH_3$  &  7(1,7)-6(0,6) EE                          &  0.51& Lovas    &         \nl
147072.7& $\rm  CH_3CN$     &  8(6)-7(6)                                 &  0.26& JPL      & NDT     \nl
147103.9& $\rm  CH_3CN   $  &  8(5)-7(5)                                 &  0.54& Lovas    &         \nl
147112.9& $\rm  U        $  &                                            &  0.16&          &         \nl
147129.2& $\rm  CH_3CN   $  &  8(4)-7(4)                                 &  0.56& Lovas    &         \nl
147149.1& $\rm  CH_3CN   $  &  8(3)-7(3)                                 &  1.28& Lovas    &         \nl
147163.3& $\rm  CH_3CN   $  &  8(2)-7(2)                                 &  1.16& Lovas    &         \nl
147171.8& $\rm  CH_3CN   $  &  8(1)-7(1)                                 &  1.53& Lovas    &         \nl
147174.6& $\rm  CH_3CN   $  &  8(0)-7(0)                                 &  1.68& Lovas    &         \nl
147206.8& $\rm  CH_3OCH_3$  &  6(3,4)-6(2,5) EE                          &  0.25& JPL      & NDT     \nl
147759.8& $\rm  U        $  &                                            &  0.34&          &         \nl
147943.7& $\rm  U        $  &                                            &  0.30&          &         \nl
147979.7& $\rm  U        $  &                                            &  0.14&          &         \nl
148028.1& $\rm HCOOCH_3  $  & 12(6,6)-11(6,5) E                          &  0.24& JPL      & NDT     \nl 
148040.6& $\rm HCOOCH_3  $  & 12(6,7)-11(6,6) E                          &  0.33& JPL      & NDT     \nl
148045.8& $\rm  HCOOCH_3 $  & 12(6,6)-11(6,5) A                          &  0.24& JPL      & NDT     \nl
148111.9& $\rm  CH_3OH   $  & $v_t=0$, 15(0)-15(-1) E                    &  0.67& XL       & NDT     \nl
148249.2& $\rm  U        $  &                                            &  0.16&          &         \nl
148500.4& $\rm CH_3OCH_3 $  & 8(3,6)-8(2,7) EE                           &  0.18& JPL      & NDT     \nl  
148503.8& $\rm CH_3OCH_3  $ & 8(3,6)-8(2,7) AA                           &  0.19& JPL      & NDT     \nl
148797.8& $\rm HCOOCH_3  $  &12(4,9)-11(4,8) E                           &  0.16& JPL      & NDT     \nl
148806.3& $\rm  HCOOCH_3  $ &12(4,9)-11(4,8) A                           &  0.19& JPL      & NDT     \nl
149439.9& $\rm  CH_3OCH_3 $ &19(3,17)-18(4,14) EE                        &  0.15& JPL      & NDT     \nl  
149532.5& $\rm  CH_3OH    $ & $v_t=0$, 14(2)-13(3) E                     &  0.68& XL       & NDT     \nl
149569.8& $\rm  CH_3OCH_3 $ & 9(3,7)-9(2,8) EE                           &  0.21& JPL      & NDT     \nl
149878.8& $\rm CH_3OCH_3$   & 5(5,1)-6(4,2) AE                           &  0.14& JPL      & NDT     \nl
150141.6& $\rm  CH_3OH    $ & $v_t=0$, 14(0)-14(-1) E                    &  0.67& Lovas    &         \nl
150155.3& $\rm  U         $ &                                            &  0.08&          &         \nl
150163.1& $\rm  CH_3OCH_3 $ & 25(4,21)-25(1,24)AA                        &  0.12& JPL      & NDT     \nl
150176.5& $\rm  NO        $ &  $2\Pi_{1/2}$ J=3/2-1/2 F=5/2-3/2(-+)      &  0.14& Lovas    &         \nl
150186.7& $\rm  U         $ &                                            &  0.10&          &         \nl
150381.1& $\rm  SO_2      $ &  15(5,11)-16(4,12)                         &  0.57& Lovas    &         \nl
150439.1& $\rm NO   $       & $2\Pi_{1/2}$ J=3/2-1/2 F=3/2-3/2(+-)       &  0.10& Lovas    &         \nl
150467.3& $\rm CH_3OCH_3  $ & 22(2,21)-21(3,18) EE                       &  0.13& JPL      & NDT     \nl
150498.3& $\rm  H_2CO     $ &  2(1,1)-1(1,0)                             &  5.47& Lovas    &         \nl
150546.5& $\rm NO     $     & $2\Pi_{1/2}$ J=3/2-1/2 F=5/2-3/2(+-)       &  0.17& Lovas    &         \nl
150600.8& $\rm HCOOCH_3$    &  12(4,8)-11(4,7) E                         &  0.28& Lovas    &         \nl
150618.3& $\rm HCOOCH_3$    &  12(4,8)-11(4,7) A                         &  0.25& Lovas    &         \nl
\enddata			      
\tablenotetext{a}{peak antenna temperatures}
\tablenotetext{b}{JPL- Pickett et al. (1998); Lovas - Lovas (1992);
Kawaguchi - Kawaguchi (1999); XL - Xu \& Lovas (1997); 
ALH -  Anderson et al. (1992); TUTT - Tsunekawa, Ukai, Toyama, \& Takagi 2000}
\tablenotetext{c}{NDT - Newly Detected Transition from the known molecule }
\end{deluxetable}
\clearpage

\begin{deluxetable}{clcrccrrrl}
\tablecaption{ Transitions of $\rm CH_3OH$}
\tablehead{
        \colhead{Frequency}                        & \colhead{Transition}               &
        \colhead{$\rm V_{LSR}$}                    & \colhead{$\rm \Delta V_{FWHM}$}    &
        \colhead{$\rm g_{I}$}                      & \colhead{$\rm g_{k}$}              & 
        \colhead{$\rm E_u/k$\tablenotemark{a}}     &\colhead{$\rm S\mu^2$\tablenotemark{a}} & 
        \colhead{$\int T_A^* dv\pm \sigma$}        & \colhead{Log L}                      \nl
        \colhead{(MHz)}                            & \colhead{$\rm v_t~ J(K_{-1}, K_{+1})$}  &
        \colhead{($\rm km~s^{-1}$)}                & \colhead{($\rm km~s^{-1}$)}        &
        \colhead{}                                 & \colhead{}                         & 
        \colhead{(K)}                              &\colhead{(Debye$^2$)}               & 
        \colhead{($\rm K~km~s^{-1}$)}              & \colhead{}
        }
\startdata                                  
139903.0 &  unassigned                  &10.3 &     7.3 & 1 &  2  & \nodata   & \nodata &  $   0.7 \pm  0.2 $ &\nodata\nl
140033.1 &0 23(-2)-23(1) E              &7.2  &     4.4 & 1 &  2  &    668.7  & 19.21   &  $   0.8 \pm  0.2 $ & 10.80 \nl
140151.1 &0 18(0)-18(-1) E              &7.2  &     5.7 & 1 &  2  &    407.4  &  6.22   &  $   1.7 \pm  0.7 $ & 11.62 \nl
140166.6 &  unassigned                  &7.8  &     4.1 & 1 &  2  & \nodata   & \nodata &  $   0.4 \pm  0.1 $ &\nodata\nl
141441.3 &1 0(0)-1(1) E                 &7.4  &     4.7 & 1 &  2  &    300.8  &  0.89   &  $   1.4 \pm  0.1 $ & 12.38 \nl
142054.4 &  unassigned                  &9.1  &     4.7 & 1 &  2  & \nodata   & \nodata &  $   0.8 \pm  0.2 $ &\nodata\nl
143108.3 &0 17(0)-17(-1) E              &7.2  &     7.4 & 1 &  2  &    367.1  &  6.61   &  $   1.7 \pm  0.3 $ & 11.59 \nl
143169.5 &0 7(3)-8(2) E                 &7.8  &     3.9 & 1 &  2  &    112.8  &  1.81   &  $   2.4 \pm  0.4 $ & 12.28 \nl
143865.8 &0 3(1)-2(1) A                 &8.0  &     4.5 & 1 &  2  &     28.4  &  2.09   &  $   3.6 \pm  0.2 $ & 12.41 \nl
144572.0 &2 3(0)-2(0) A                 &6.9  &     6.2 & 1 &  2  &    522.1  &  2.35   &  $   0.6 \pm  0.1 $ & 11.59 \nl
144583.9 &2 3(-1)-2(-1) E               &8.1  &     7.3 & 1 &  2  &    545.6  &  2.09   &  $   0.7 \pm  0.1 $ & 11.67 \nl
144589.8 &1 3(1)-2(1) A                 &7.7  &     8.4 & 1 &  2  &    339.4  &  2.16   &  $   1.2 \pm  0.2 $ & 11.92 \nl
144728.7 &1 3(-2)-2(-2) E \&            &6.8  &     7.7 & 1 &  2  &    378.8  &  1.34   &  $   2.5 \pm  0.2 $ & 12.44 \nl
         &    3(2)-2(2) A               &     &         &   &     &           &         &                     &       \nl
144734.5 &1 3(1)-2(1) E                 &6.4  &    10.4 & 1 &  2  &    304.4  &  2.15   &  $   4.5 \pm  0.3 $ & 12.50 \nl
144750.2 &1 3(-1)-2(-1) E               &6.2  &     7.7 & 1 &  2  &    426.2  &  2.15   &  $   1.5 \pm  0.2 $ & 12.03 \nl
144768.2 &1 3(0)-2(1) A                 &7.7  &     5.6 & 1 &  2  &    437.9  &  2.42   &  $   1.1 \pm  0.1 $ & 11.83 \nl
144878.6 &1 3(1)-2(1) A                 &7.4  &     4.0 & 1 &  2  &     26.1  &  2.16   &  $   0.9 \pm  0.1 $ & 11.76 \nl
145093.6 &0 3(0)-2(0) E                 &8.0  &     5.0 & 1 &  2  &     27.1  &  2.42   &  $   9.8 \pm  0.2 $ & 12.79 \nl
145097.5 &0 3(-1)-2(-1) E               &8.7  &     6.7 & 1 &  2  &     19.5  &  2.16   &  $  13.1 \pm  0.3 $ & 12.97 \nl
145103.2 &0 3(0)-2(0) A                 &8.5  &     5.8 & 1 &  2  &     13.9  &  2.42   &  $  11.0 \pm  0.8 $ & 12.84 \nl
145124.4 &0 3(2)-2(2) A                 &5.6  &     8.0 & 1 &  2  &     51.7  &  1.36   &  $  16.1 \pm  0.2 $ & 13.24 \nl
145126.4 &0 3(2)-2(2) E \&              &9.7  &     8.0 & 1 &  2  &     39.9  &  1.34   &  $  16.1 \pm  0.2 $ & 13.26 \nl
         &    3(-2)-3(-2)E              &     &         &   &     &           &         &                     &       \nl
145131.9 &0 3(1)-2(1) E                 &7.2  &     7.7 & 1 &  2  &     35.0  &  2.21   &  $  14.4 \pm  0.3 $ & 13.01 \nl
145133.5 &0 3(2)-2(2) A                 &10.5 &     7.8 & 1 &  2  &     38.0  &  1.35   &  $  14.6 \pm  0.2 $ & 13.20 \nl
145766.1 &0 16(0)-16(-1) E              &7.2  &     6.5 & 1 &  2  &    327.9  &  6.96   &  $   1.9 \pm  0.3 $ & 11.60 \nl
146368.3 &0 3(1)-2(1)A                  &8.2  &     4.2 & 1 &  2  &     28.6  &  2.16   &  $   4.3 \pm  0.2 $ & 12.47 \nl
146618.8 &0 9(0)-8(1) A                 &9.1  &     8.4 & 1 &  2  &    104.5  &  8.33   &  $   9.9 \pm  1.2 $ & 12.26 \nl
148111.9 &0 15(0)-15(-1) E              &7.5  &     7.0 & 1 &  2  &    289.8  &  7.26   &  $   4.9 \pm  0.2 $ & 12.05 \nl
149532.5 &0 14(2)-13(3) E               &7.4  &     6.2 & 1 &  2  &    265.2  &  4.20   &  $   4.0 \pm  0.3 $ & 12.14 \nl
150141.6 &0 14(0)-14(-1) E              &8.0  &     6.9 & 1 &  2  &    255.2  &  7.49   &  $   4.5 \pm  0.3 $ & 11.93 \nl
\enddata            
\tablenotetext{a}{From Xu \& Lovas (1997) except for 3(0)-2(0) A transition 
(144572.0 MHz) from  Anderson et al. (1990) and 3(-1)-2(-1) E transition (144583.9 MHz) from Anderson et al. (1992) 
}
\end{deluxetable}                                          
\clearpage

\begin{deluxetable}{clcrccrrrl}
\tablecaption{ Transitions of $\rm ^{13}CH_3OH$}
\tablehead{
        \colhead{Frequency}                        & \colhead{Transition}               &
        \colhead{$\rm V_{LSR}$}                    & \colhead{$\rm \Delta V_{FWHM}$}    &
        \colhead{$\rm g_{I}$}                      & \colhead{$\rm g_{k}$}              & 
        \colhead{$\rm E_u/k$\tablenotemark{a}}     &\colhead{$\rm S\mu^2$\tablenotemark{a}} & 
        \colhead{$\int T_A^* dv\pm \sigma$}        & \colhead{Log L}                      \nl
        \colhead{(MHz)}                            & \colhead{$\rm v_t~ J(K_{-1}, K_{+1})$}  &
        \colhead{($\rm km~s^{-1}$)}                & \colhead{($\rm km~s^{-1}$)}        &
        \colhead{}                                 & \colhead{}                         & 
        \colhead{(K)}                              &\colhead{(Debye$^2$)}               &
        \colhead{($\rm K~km~s^{-1}$)}              & \colhead{}
        }
\startdata
141597.1 &0 3(-1)-2(-1) E               & 9.3 &  7.0    & 1 &  2  &   26.7    &  8.62   &  $  2.0  \pm 0.2  $ &11.55  \nl
141602.5 &0 3(0)-2(0) A                 & 8.2 &  3.1    & 1 &  2  &   13.6    &  4.72   &  $  0.8  \pm 0.1  $ &11.41  \nl
141623.5 &0 3(2)-2(2) E \&              & 7.8 &  6.4    & 1 &  2  &   39.6    & 22.80   &  $  2.0  \pm 0.2  $ &11.12  \nl
         &0  3(-2)-2(-2) E              &     &         &   &     &           &         &                     &       \nl
141629.2 &0 3(1)-2(1) E                 & 6.7 &  6.9    & 1 &  2  &   34.7    & 19.36   &  $  2.0  \pm 0.2  $ &11.19  \nl
\enddata            
\tablenotetext{a}{From Xu \& Lovas (1997)
}
\end{deluxetable}                                          
\clearpage      

\begin{deluxetable}{clrrccrrrr}
\tablecaption{ Transitions of $\rm CH_3OCH_3$}
\tablehead{
        \colhead{Frequency}                        & \colhead{Transition}               &
        \colhead{$\rm V_{LSR}$}                    & \colhead{$\rm \Delta V_{FWHM}$}    &
        \colhead{$\rm g_{I}$}                      & \colhead{$\rm g_{k}$}              & 
        \colhead{$\rm E_u/k$\tablenotemark{a}}     &\colhead{$\rm S\mu^2$\tablenotemark{a}} & 
        \colhead{$\int T_A^* dv\pm \sigma$}        & \colhead{Log L}                      \nl
        \colhead{(MHz)}                            & \colhead{$\rm J(K_{-1}, K_{+1})$}  &
        \colhead{($\rm km~s^{-1}$)}                & \colhead{($\rm km~s^{-1}$)}        &
        \colhead{}                                 & \colhead{}                         & 
        \colhead{(K)}                              &\colhead{(Debye$^2$)}               &
        \colhead{($\rm K~km~s^{-1}$)}              & \colhead{}
        }
\startdata
139500.4 & 9(3,6)-9(2,7) AE    &  7.1& 5.9 &  1 & 0.40 &   53.7 &  1.37  &$ 0.7 \pm0.1 $&  12.56\nl
139503.7 & 9(3,6)-9(2,7) EE    &  7.5& 5.4 &  1 & 1.60 &   53.7 &  3.65  &$ 0.9 \pm0.2 $&  11.68\nl
139506.9 & 9(3,6)-9(2,7) AA    &  9.0& 5.3 &  1 & 1.00 &   53.7 &  2.28  &$ 0.9 \pm0.1 $&  12.07\nl
141829.1 & 8(3,5)-8(2,6) EA    &  7.9& 3.8 &  1 & 0.67 &   45.5 &  0.78  &$ 0.7 \pm0.2 $&  12.57\nl
141832.3 & 8(3,5)-8(2,6) EE    &  8.4& 3.7 &  1 & 2.67 &   45.5 &  3.14  &$ 1.4 \pm0.2 $&  11.70\nl
141835.5 & 8(3,5)-8(2,6) AA    &  8.1& 4.4 &  1 & 1.00 &   45.5 &  1.18  &$ 0.8 \pm0.5 $&  12.32\nl
143599.4 &  7(3,4)-7(2,5) AE   &  7.8& 4.5 &  1 & 0.40 &   38.2 &  0.99  &$ 0.7 \pm0.1 $&  12.74\nl
143603.0 &  7(3,4)-7(2,5) EE   &  8.1& 2.1 &  1 & 1.60 &   38.2 &  2.65  &$ 0.6 \pm0.1 $&  11.61\nl
143606.2 &  7(3,4)-7(2,5) AA   &  7.7& 2.9 &  1 & 1.00 &   38.2 &  1.66  &$ 0.5 \pm0.1 $&  11.98\nl
144859.0 &  6(3,3)-6(2,4) EE   &  7.6& 4.1 &  1 & 2.67 &   31.8 &  2.16  &$ 0.7 \pm0.1 $&  11.58\nl
145680.4 &  5(3,2)-5(2,3) EE   &  8.4& 4.7 &  1 & 1.60 &   26.3 &  1.60  &$ 0.7 \pm0.2 $&  11.89\nl
147024.9 &  7(1,7)-6(0,6) EE   &  7.9& 6.0 &  1 & 2.67 &   26.1 &  3.38  &$ 3.0 \pm0.5 $&  11.98\nl
147206.8 &  6(3,4)-6(2,5) EE   &  8.5& 5.0 &  1 & 1.60 &   31.8 &  2.14  &$ 1.2 \pm0.3 $&  12.01\nl
148500.4 & 8(3,6)-8(2,7) EE    &  7.5& 4.9 &  1 & 1.60 &   45.5 &  3.04  &$ 1.0 \pm0.1 $&  11.75\nl  
148503.8 & 8(3,6)-8(2,7) AA    &  7.0& 3.8 &  1 & 1.00 &   45.5 &  1.90  &$ 0.7 \pm0.1 $&  12.04\nl  
149439.9 &19(3,17)-18(4,14) EE &  9.8& 6.2 &  1 & 2.67 &  185.9 &  1.95  &$ 1.0 \pm0.2 $&  11.72\nl  
149569.8 & 9(3,7)-9(2,8) EE    &  7.9& 3.9 &  1 & 2.67 &   53.7 &  3.47  &$ 0.9 \pm0.2 $&  11.45\nl
149878.8 & 5(5,1)-6(4,2) AE    &  9.6& 5.9 &  1 & 0.33 &   48.8 &  0.01  &$ 0.9 \pm0.2 $&  14.88\nl
150163.1 &25(4,21)-25(1,24) AA &  7.3& 8.0 &  1 & 1.00 &  321.0 &  8.81  &$ 1.0 \pm0.2 $&  11.50\nl
150467.3 &22(2,21)-21(3,18) EE &  6.9& 6.6 &  1 & 1.60 &  233.1 &  0.67  &$ 0.9 \pm0.2 $&  12.40\nl
\enddata                                                       
\tablenotetext{a}{From JPL catalog}
\end{deluxetable}
\clearpage

\begin{deluxetable}{clrrccrrrr}
\tablecaption{ Transitions of $\rm HCOOCH_3$}
\tablehead{
        \colhead{Frequency}                        & \colhead{Transition}               &
        \colhead{$\rm V_{LSR}$}                    & \colhead{$\rm \Delta V_{FWHM}$}    &
        \colhead{$\rm g_{I}$}                      & \colhead{$\rm g_{k}$}              & 
        \colhead{$\rm E_u/k$\tablenotemark{a}}     &\colhead{$\rm S\mu^2$\tablenotemark{a}} & 
        \colhead{$\int T_A^* dv\pm \sigma$}        & \colhead{Log L}                      \nl
        \colhead{(MHz)}                            & \colhead{$\rm J(K_{-1}, K_{+1})$}  &
        \colhead{($\rm km~s^{-1}$)}                & \colhead{($\rm km~s^{-1}$)}        &
        \colhead{}                                 & \colhead{}                         & 
        \colhead{(K)}                              &\colhead{(Debye$^2$)}               &
        \colhead{$\rm K~km~s^{-1}$}                & \colhead{}
        }
\startdata
141260.4 & 11(3,8)-10(3,7) A        &   8.4   &   4.9 &  2&  1&   45.8  & 27.13 &  $ 2.0 \pm0.2 $&   11.05 \nl
141653.0 & 11(2,9)-10(2,8) E        &   7.3   &   8.5 &  1&  2&   43.3  & 28.18 &  $ 4.4 \pm0.2 $&   11.37 \nl
141667.0 & 11(2,9)-10(2,8) A        &   7.6   &   3.9 &  2&  1&   43.3  & 28.18 &  $ 1.7 \pm0.1 $&   10.96 \nl
142735.1 & 13(1,13)-12(1,12) A      &   9.4   &   5.9 &  2&  1&   49.0  & 34.04 &  $ 2.1 \pm0.2 $&   10.96 \nl
142815.5 & 13(0,13)-12(0,12) A      &   6.6   &   7.9 &  2&  1&   49.0  & 34.04 &  $ 1.5 \pm0.3 $&   10.82 \nl
142924.5 & 13(1,13)-12(0,12) E      &   7.0   &   9.0 &  1&  2&   49.0  & 30.31 &  $ 1.5 \pm0.4 $&   10.87 \nl
143240.5 &  12(1,11)-11(1,10) A     &   9.0   &   4.9 &  2&  1&   44.4  & 30.85 &  $ 1.4 \pm0.4 $&   10.83 \nl
146977.7 &  12(3,10)-11(3,9) E      &   9.1   &   6.0 &  1&  2&   52.1  & 29.78 &  $ 3.9 \pm0.4 $&   11.28 \nl
146988.0 &  12(3,10)-11(3,9) A      &   8.0   &   2.9 &  2&  1&   52.1  & 29.78 &  $ 1.4 \pm0.2 $&   10.83 \nl
148028.1 & 12(6,6)-11(6,5) E        &   8.1   &   3.9 &  1&  2&   70.0  & 23.93 &  $ 1.1 \pm0.1 $&   10.82 \nl
148040.6 & 12(6,7)-11(6,6) E        &   9.3   &   5.4 &  1&  2&   70.0  & 23.93 &  $ 1.8 \pm0.2 $&   11.03 \nl
148045.8 & 12(6,6)-11(6,5) A        &   8.5   &   4.1 &  2&  1&   70.0  & 23.93 &  $ 0.9 \pm0.1 $&   10.75 \nl
148797.8 &12(4,9)-11(4,8) E         &   8.0   &   3.2 &  1&  2&   56.9  & 28.32 &  $ 0.7 \pm0.2 $&   10.52 \nl
148806.3 &12(4,9)-11(4,8) A         &   8.4   &   3.5 &  2&  1&   56.9  & 28.34 &  $ 0.8 \pm0.2 $&   10.60 \nl
150600.8 &  12(4,8)-11(4,7) E       &   8.1   &   3.3 &  1&  2&   57.1  & 28.33 &  $ 1.0 \pm0.2 $&   10.70 \nl
150618.3 &  12(4,8)-11(4,7) A       &   8.6   &   3.4 &  2&  1&   57.1  & 28.35 &  $ 1.1 \pm0.1 $&   10.74 \nl
\enddata					       		 
\tablenotetext{a}{From Bauder (1979) for all $E_u/k$ and $S\mu^2$ except for those 
for J=13$-$12 transitions which are from Plummer et al (1984).}
 
\end{deluxetable}
\clearpage

\begin{deluxetable}{clrrccrrrr}
\tablecaption{ Transitions of $\rm C_2H_5CN$}
\tablehead{
        \colhead{Frequency}                        & \colhead{Transition}               &
        \colhead{$\rm V_{LSR}$}                    & \colhead{$\rm \Delta V_{FWHM}$}    &
        \colhead{$\rm g_{I}$}                      & \colhead{$\rm g_{k}$}              & 
        \colhead{$\rm E_u/k$\tablenotemark{a}}     &\colhead{$\rm S\mu^2$\tablenotemark{a}} & 
        \colhead{$\int T_A^* dv\pm \sigma$}        & \colhead{Log L}                      \nl
        \colhead{(MHz)}                            & \colhead{$\rm J(K_{-1}, K_{+1})$}  &
        \colhead{($\rm km~s^{-1}$)}                & \colhead{($\rm km~s^{-1}$)}        &
        \colhead{}                                 & \colhead{}                         & 
        \colhead{(K)}                              &\colhead{(Debye$^2$)}               &
        \colhead{($\rm K~km~s^{-1}$)}              & \colhead{}
        }
\startdata
138351.1& 16(1,16)-15(1,15)            & 4.4&12.7 & 1 &   1 &  57.8  & 235.7 & $  3.7\pm 0.3$ & 10.68\nl
139335.9& 16(0,16)-15(0,15)            & 3.0&18.8 & 1 &   1 &  57.5  & 236.1 & $  2.5\pm 0.2$ & 10.51\nl 
142346.3& 16(2,15)-15(2,14)            & 3.9&15.1 & 1 &   1 &  62.7  & 233.0 & $  3.3\pm 0.4$ & 10.63\nl
143337.7& 16(7,9)-15(7,8) \&           & 6.0&12.3 & 1 &   1 & 113.0  & 191.5 & $  4.1\pm 0.3$ & 10.79\nl
        &     16(7,10)-15(7,9)         &    &     &   &     &        &       &                &      \nl
143343.9& 16(9,8)-15(9,7) \&           & 7.5&11.5 & 1 &   1 & 148.6  & 161.9 & $  2.3\pm 0.8$ & 10.83\nl
        &     16(9,7)-15(9,6)          &    &     &   &     &        &       &                &      \nl
143357.2& 16(6,10)-15(6,9) \&          & 4.8& 9.7 & 1 &   1 &  98.6  & 203.6 & $  2.8\pm 0.3$ & 10.61\nl
        &  16(6,11)-15(6,10)           &    &     &   &     &        &       &                &      \nl
143360.4& 16(10,7)-15(10,6) \&         & 6.1&11.0 & 1 &   1 & 169.6  & 144.3 & $  2.5\pm 0.8$ & 10.72\nl
        &    16(10,6)-15(10,5)         &    &     &   &     &        &       &                &      \nl
143383.0& 16(11,6)-15(11,5) \&         & 4.2&11.0 & 1 &   1 & 192.9  & 124.9 & $  1.2\pm 0.2$ & 10.47\nl
        &    16(11,5)-15(11,4)         &    &     &   &     &        &       &                &      \nl
143407.2& 16(5,12)-15(5,11) \&         & 5.8& 8.2 & 1 &   1 &  86.3  & 213.8 & $  2.9\pm 0.4$ & 10.60\nl
        &    16(5,11)-15(5,10)         &    &     &   &     &        &       &                &      \nl
143410.8& 16(12,5)-15(12,4) \&         & 8.6&11.7 & 1 &   1 & 218.4  & 103.6 & $  2.8\pm 1.0$ & 10.91\nl
        & 16(12,4)-15(12,3)            &    &     &   &     &        &       &                &      \nl
143507.0& 16(4,13)-15(4,12)            & 5.5&16.6 & 1 &   1 &  76.4  & 222.0 & $  3.4\pm 0.3$ & 10.66\nl
143529.2&  16(3,14)-15(3,13)           & 5.9&12.1 & 1 &   1 &  68.6  & 228.5 & $  3.4\pm 0.4$ & 10.65\nl
143535.3&  16(4,12)-15(4,11)           & 8.2&16.0 & 1 &   1 &  76.4  & 222.1 & $  3.5\pm 1.1$ & 10.86\nl
146120.0& 16(2,14)-15(2,13)            & 1.9&13.9 & 1 &   1 &  63.7  & 233.2 & $  3.7\pm 0.5$ & 10.67\nl
146894.5& 17(1,17)-16(1,16)            & 3.7&12.2 & 1 &   1 &  64.8  & 250.6 & $  4.1\pm 0.5$ & 10.68\nl
\enddata                              
\tablenotetext{a}{ From JPL catalog
}
\end{deluxetable}
\clearpage

\begin{deluxetable}{clrrccrrrr}
\tablecaption{ Transitions of $\rm SO_2$}
\tablehead{
        \colhead{Frequency}                        & \colhead{Transition}               &
        \colhead{$\rm V_{LSR}$}                    & \colhead{$\rm \Delta V_{FWHM}$}    &
        \colhead{$\rm g_{I}$}                      & \colhead{$\rm g_{k}$}              & 
        \colhead{$\rm E_u/k$\tablenotemark{a}}     &\colhead{$\rm S\mu^2$\tablenotemark{a}} & 
        \colhead{$\int T_A^* dv\pm \sigma$}        & \colhead{Log L}                      \nl
        \colhead{(MHz)}                            & \colhead{$\rm J(K_{-1}, K_{+1})$}  &
        \colhead{($\rm km~s^{-1}$)}                & \colhead{($\rm km~s^{-1}$)}        &
        \colhead{}                                 & \colhead{}                         &
        \colhead{(K)}                              &\colhead{(Debye$^2$)}               &
        \colhead{($\rm K~km~s^{-1}$)}              & \colhead{}
        }
\startdata
139355.1 & 5(3,3)-6(2,4)      & 8.3    &13.9 & 1 & 1 &    35.9 &   1.3  &$ 3.8\pm 0.3$ &  12.93 \nl
         &                    &18.0    &14.5 & 1 & 1 &    35.9 &   1.3  &$ 2.2\pm 0.7$ &  12.70 \nl
139474.5 & 26(7,19)-27(6,22)  & 6.7    & 9.6 & 1 & 1 &   443.4 &  10.2  &$ 0.9\pm 0.2$ &  11.43 \nl
140306.2 & 6(2,4)-6(1,5)      & 7.4    &15.3 & 1 & 1 &    29.2 &  10.2  &$19.3\pm 0.5$ &  12.76 \nl	    
         &                    &14.8    &22.0 & 1 & 1 &    29.2 &  10.2  &$17.1\pm 3.0$ &  12.71 \nl	    
143057.1 & 16(2,14)-16(1,15)  & 6.9    &17.8 & 1 & 1 &   137.6 &  34.6  &$18.0\pm 1.4$ &  12.19 \nl
         &                    &17.0    &19.2 & 1 & 1 &   137.6 &  34.6  &$11.6\pm 1.6$ &  12.00 \nl
146550.1 &  10(4,6)-11(3,9)   & 5.5    &30.5 & 1 & 1 &    89.9 &   3.4  &$10.2\pm 0.5$ &  12.95 \nl
146605.5 &  4(2,2)-4(1,3)     & 5.9    &24.3 & 1 & 1 &    19.0 &   6.1  &$25.2\pm 1.5$ &  13.08 \nl
150381.1 &  15(5,11)-16(4,12) & 6.8    &21.4 & 1 & 1 &   171.8 &   5.4  &$10.6\pm 0.8$ &  12.75 \nl
         &                    &10.1    &34.2 & 1 & 1 &   171.8 &   5.4  &$10.4\pm 1.2$ &  12.73 \nl
 
\enddata
\tablenotetext{a}{From Helminger \& DeLucia (1985) }
\end{deluxetable}
\clearpage

\begin{deluxetable}{clrrccrrrr}
\tablecaption{ Transitions of $\rm CH_3CN$}
\tablehead{
        \colhead{Frequency}                        & \colhead{Transition}               &
        \colhead{$\rm V_{LSR}$}                    & \colhead{$\rm \Delta V_{FWHM}$}    &
        \colhead{$\rm g_{I}$}                      & \colhead{$\rm g_{k}$}              & 
        \colhead{$\rm E_u/k$\tablenotemark{a}}     &\colhead{$\rm S\mu^2$\tablenotemark{a}} & 
        \colhead{$\int T_A^* dv\pm \sigma$}        & \colhead{Log L}                      \nl
        \colhead{(MHz)}                            & \colhead{$\rm J(K_{-1}, K_{+1})$}  &
        \colhead{($\rm km~s^{-1}$)}                & \colhead{($\rm km~s^{-1}$)}        &
        \colhead{}                                 & \colhead{}                         & 
        \colhead{(K)}                              &\colhead{(Debye$^2$)}               &
        \colhead{($\rm K~km~s^{-1}$)}              & \colhead{}
        }
\startdata
147072.7 & 8(6)-7(6)  &  3.7     &  13.0  & 0.50 &  2.0  &   275.3   &   53.6 &  $  3.4\pm0.5  $ & 11.27 \nl
147103.9 & 8(5)-7(5)  &  7.9     &  13.4  & 0.25 &  2.0  &   209.6   &   74.6 &  $  5.9\pm0.6  $ & 11.66 \nl    
147129.2 & 8(4)-7(4)  &  7.5     &   8.7  & 0.25 &  2.0  &   145.6   &   91.9 &  $  4.5\pm0.6  $ & 11.46 \nl
147149.1 & 8(3)-7(3)  &  7.8     &   5.1  & 0.50 &  2.0  &    95.8   &  105.3 &  $  6.8\pm0.4  $ & 11.27 \nl
147163.3 & 8(2)-7(2)  &  6.9     &   8.0  & 0.25 &  2.0  &    60.3   &  114.8 &  $  7.6\pm0.9  $ & 11.59 \nl
147171.8 & 8(1)-7(1)  &  6.9     &   8.9  & 0.25 &  2.0  &    38.9   &  120.6 &  $ 13.5\pm1.6  $ & 11.81 \nl
147174.6 & 8(0)-7(0)  &  8.4     &   9.7  & 0.50 &  1.0  &    31.8   &  122.5 &  $ 15.6\pm1.8  $ & 11.87 \nl
\enddata                              
\tablenotetext{a}{From Pearson \& M\"uller (1996)
}
 
\end{deluxetable}
\clearpage

\begin{deluxetable}{clrrccrrrr}
\tablecaption{ Transitions of $\rm H_2CO$}		       
\tablehead{
        \colhead{Frequency}                        & \colhead{Transition}               &
        \colhead{$\rm V_{LSR}$}                    & \colhead{$\rm \Delta V_{FWHM}$}    &
        \colhead{$\rm g_{I}$}                      & \colhead{$\rm g_{k}$}              & 
        \colhead{$\rm E_u/k$\tablenotemark{a}}     &\colhead{$\rm S\mu^2$\tablenotemark{a}} & 
        \colhead{$\int T_A^* dv\pm \sigma$}        & \colhead{Log L}                      \nl
        \colhead{(MHz)}                            & \colhead{$\rm J(K_{-1}, K_{+1})$}  &
        \colhead{($\rm km~s^{-1}$)}                & \colhead{($\rm km~s^{-1}$)}        &
        \colhead{}                                 & \colhead{}                         & 
        \colhead{(K)}                              &\colhead{(Debye$^2$)}               &
        \colhead{($\rm K~km~s^{-1}$)}              & \colhead{}
        }
\startdata
140839.5 & 2(1,2)-1(1,1) &  8.6  &     5.5 & 0.75&1.0 &  21.9 &    8.2  &     $  17.7\pm0.3$  &  13.24        \nl
         &               &  8.3  &    19.9 & 0.75&1.0 &  21.9 &    8.2  &     $   7.4\pm0.6$  &  12.86        \nl
145602.9 & 2(0,2)-1(0,1) &  8.9  &     4.7 & 0.25&1.0 &  10.5 &   10.9  &     $  12.4\pm0.2$  &  13.43        \nl
         &               &  7.0  &    15.1 & 0.25&1.0 &  10.5 &   10.9  &     $   4.2\pm0.7$  &  12.95        \nl
150498.3 & 2(1,1)-1(1,0) &  8.8  &     5.7 & 0.75&1.0 &  22.6 &    8.2  &     $  33.3\pm0.3$  &  13.49        \nl
         &               &  7.8  &    17.3 & 0.75&1.0 &  22.6 &    8.2  &     $  20.4\pm0.7$  &  13.28        \nl
\enddata
\tablenotetext{a}{Winnewisser et al. (1979)
}

\end{deluxetable}
\clearpage

\begin{deluxetable}{clrrrrrc}
\tablewidth{40pc}
\tablecaption{ Physical Quantities}
\tablehead{
        \colhead{Molecule}  & \colhead{$\rm V_{LSR}$\tablenotemark{a}} &
        \colhead{$\rm \Delta V_{FWHM}$\tablenotemark{a}} & \colhead{$\rm T_{rot}$\tablenotemark{b}}  & 
        \colhead{N\tablenotemark{b}} & \colhead{$\rm T_{rot}$\tablenotemark{c}}  & 
        \colhead{N\tablenotemark{c}}& \colhead{References\tablenotemark{d}} \nl 
        \colhead{        }  & \colhead{($\rm km~s^{-1}$)} &
        \colhead{($\rm km~s^{-1}$)} & \colhead{(K)}  & 
        \colhead{($\rm cm^{-2}$)}   & \colhead{(K)}  & 
        \colhead{($\rm cm^{-2}$)}   & \colhead{                           }    
        }
\startdata
$\rm  C_2H_5CN   $ &$  5.3\pm1.8$&$ 12.9\pm 2.7 $&    &    \nodata        &       &        \nodata        &         \nl
$\rm  CH_3CN     $ &$  7.0\pm1.4$&$  9.5\pm 2.7 $& 273&$1.3\times 10^{15}$& 77-274&$0.6-6.5\times 10^{16}$&2,4,5,6  \nl
$\rm  CH_3OCH_3  $ &$  8.0\pm0.8$&$  4.9\pm 1.4 $&    &     \nodata       &       &        \nodata        &         \nl
$\rm  H_2CO      $ &$  8.8\pm0.1$&$  5.3\pm 0.5 $&  10&$1.6\times 10^{14}$&   40  &$    2.4\times 10^{14}$& 5       \nl
                   &$  7.7\pm0.7$&$ 17.4\pm 2.4 $&    &    \nodata        &       &        \nodata        &         \nl
$\rm  HCOOCH_3   $ &$  8.2\pm0.8$&$  5.1\pm 1.9 $&  22&$2.3\times 10^{15}$& 22-90 &$0.3-3.0\times 10^{15}$& 4,5,6   \nl
                   &             &               &  18&$2.6\times 10^{14}$&       &                       &         \nl
$\rm  SO_2       $ &$  6.8\pm0.9$&$ 19.0\pm 7.0 $& 120&$1.7\times 10^{16}$&106-150&$1.5-3.0\times 10^{16}$&1,4,5,6  \nl
                   &$ 15.0\pm3.5$&$ 22.5\pm 8.4 $&    &    \nodata        &       &        \nodata        &         \nl
$\rm  NO         $ &$  8.2\pm0.4$&$  9.9\pm 0.7 $&    &    \nodata        &       &        \nodata        &         \nl
$\rm  CH_3OH     $ &$  7.8\pm1.1$&$  6.4\pm 1.6 $& 158&$1-2\times 10^{16}$&114-192&$0.8-30 \times 10^{16}$&3,5,6,7  \nl
$\rm^{13}CH_3OH  $ &$  8.0\pm0.9$&$  5.9\pm 1.6 $&  40&$1.2\times 10^{14}$&130-180&$4.2-50 \times 10^{14}$&  7      \nl
$\rm  CS         $ &$  8.3\pm0.0$&$  5.7\pm 0.1 $&  65&$2.0\times 10^{14}$&       &        \nodata        &         \nl
                   &$  6.3\pm0.5$&$ 24.1\pm 1.6 $&    &    \nodata        &       &        \nodata        &         \nl
$\rm^{13}CS      $ &$  9.0\pm0.2$&$  5.3\pm 0.4 $&    &    \nodata        &       &        \nodata        &         \nl
$\rm  C^{34}S    $ &$  9.1\pm0.1$&$  5.2\pm 0.2 $&  65&$1.1\times 10^{13}$&       &        \nodata        &         \nl
$\rm  C^{33}S    $ &$  8.1\pm0.4$&$  5.3\pm 0.9 $&    &    \nodata        &       &        \nodata        &         \nl
$\rm  H_2CCO     $ &$  7.8\pm0.5$&$  4.1\pm 0.0 $&    &    \nodata        &       &        \nodata        &         \nl
$\rm  H_2CS      $ &$  8.7\pm0.1$&$  5.7\pm 0.4 $&    &    \nodata        &       &        \nodata        &         \nl
$\rm  C_2H_5OH   $ &$  9.8\pm0.8$&$  7.3\pm 2.4 $&    &    \nodata        &       &        \nodata        &         \nl
$\rm  CH_2CHCN   $ &$  9.3\pm0.3$&$  2.1\pm 0.2 $&    &    \nodata        &       &        \nodata        &         \nl
$\rm  DCN        $ &$  8.3\pm0.1$&$  7.5\pm 0.2 $&    &    \nodata        &       &        \nodata        &         \nl
$\rm  HC_3N      $ &$  8.7\pm0.1$&$  5.8\pm 0.3 $&    &    \nodata        &       &        \nodata        &         \nl
                   &$  5.6\pm0.8$&$ 17.2\pm 1.8 $&    &    \nodata        &       &        \nodata        &         \nl
$\rm  OCS        $ &$  7.8\pm0.3$&$  6.5\pm 0.8 $&    &    \nodata        &       &        \nodata        &         \nl
                   &$  5.5\pm0.6$&$ 15.8\pm 1.2 $&    &    \nodata        &       &        \nodata        &         \nl
\enddata 
\tablenotetext{a}{These were obtained from the Gaussian fit of the spectra. 
In case of molecules with multi line detections the average values are given.}
\tablenotetext{b}{This study. $\rm T_{ex}$ for CS was adopted from Shutton et al. (1995)  }
\tablenotetext{c}{Other studies}
\tablenotetext{d}{1-Schloerb et al (1983), 2-Johanson et al. (1984), 
3-Sutton et al. (1985), 4-Blake et al. (1986),
5-Turner (1991), 6-Ziurys \& McGonagle (1993), 7-Sutton et al. (1995) }
\end{deluxetable}
\clearpage                                                         

\begin{deluxetable}{cccrrr}
\tablewidth{40pc}
\tablecaption{ Transitions of $\rm CS$}
\tablehead{
        \colhead{Frequency }    & \colhead{Molecule}  & 
        \colhead{$\rm V_{LSR}$} &
        \colhead{$\Delta V_{FWHM}$ }   & 
        \colhead{$\rm \int T_A^* dv$}    & 
        \colhead{$\rm T_A^*$ }  \nl 
        \colhead{(MHz)}    & \colhead{}  & 
        \colhead{($\rm km~s^{-1}$)} &
        \colhead{($\rm km~s^{-1}$)}   & 
        \colhead{($\rm K~km~s^{-1}$)}    & 
        \colhead{(K)}   
        }
\startdata
138739.3& $\rm ^{13}CS $&  9.0&      5.3&  $   3.2\pm0.2$   &  0.56  \nl
144617.1& $\rm  C^{34}S$&  9.1&      5.2&  $   6.9\pm0.3$   &  1.23  \nl
145755.6& $\rm  C^{33}S$&  8.1&      5.0&  $   1.0\pm0.2$   &  0.19  \nl
146969.0& $\rm  CS     $&  8.2&      5.5&  $  37.7\pm0.5$   &  6.46  \nl
        &               &  6.5&     21.5&  $  28.4\pm1.2$   &  1.24  \nl
\enddata

\end{deluxetable}
\clearpage
 
\begin{deluxetable}{crrc}
\tablewidth{40pc}
\tablecaption{ Unidentified  lines\tablenotemark{a}}
\tablehead{
        \colhead{Frequency}      &
        \colhead{$\rm \Delta V_{FWHM}$ } & 
        \colhead{$\rm T_A^*$} & \colhead{$\int T_A^* dv$ }  \nl 
        \colhead{(MHz)}      &
        \colhead{($\rm km~s^{-1}$)} & 
        \colhead{(K)} & \colhead{($\rm K~km~s^{-1}$)} 
        }
\startdata
138259.9&      8.1  & $ 0.10\pm 0.03$ &  0.89  \nl
138343.2&      8.4  & $ 0.09\pm 0.03$ &  0.81  \nl
138396.4&      9.3  & $ 0.13\pm 0.03$ &  1.29  \nl
139256.7&     10.2  & $ 0.09\pm 0.02$ &  0.93  \nl 
139416.9&      7.6  & $ 0.10\pm 0.02$ &  0.81  \nl
139432.5&     11.1  & $ 0.14\pm 0.01$ &  1.80  \nl
139436.0&      7.7  & $ 0.14\pm 0.02$ &  1.10  \nl
139561.9&      4.8  & $ 0.32\pm 0.02$ &  1.66  \nl
139582.1&      4.9  & $ 0.13\pm 0.02$ &  0.68  \nl
139862.3&     17.5  & $ 0.08\pm 0.02$ &  1.47  \nl
139878.4&     17.2  & $ 0.10\pm 0.02$ &  1.74  \nl
140174.6&      5.4  & $ 0.08\pm 0.02$ &  0.47  \nl
141646.5&      6.3  & $ 0.13\pm 0.04$ &  0.89  \nl
141983.7&      2.8  & $ 0.36\pm 0.04$ &  1.07  \nl
143006.7&      3.5  & $ 0.20\pm 0.06$ &  0.73  \nl
143263.3&      9.5  & $ 0.07\pm 0.03$ &  0.75  \nl
143821.6&      3.8  & $ 0.20\pm 0.04$ &  0.81  \nl
144267.0&     21.3  & $ 0.11\pm 0.03$ &  2.40  \nl
144351.4&      9.0  & $ 0.14\pm 0.03$ &  1.34  \nl
144370.2&     14.6  & $ 0.13\pm 0.04$ &  2.04  \nl
145876.2&     11.8  & $ 0.15\pm 0.05$ &  1.89  \nl
146129.6&      9.0  & $ 0.17\pm 0.05$ &  1.62  \nl
146372.4&      5.4  & $ 0.42\pm 0.06$ &  2.40  \nl
146622.4&      7.3  & $ 0.74\pm 0.09$ &  5.73  \nl
146984.5&      2.0  & $ 0.38\pm 0.06$ &  0.82  \nl
147112.9&     13.1  & $ 0.16\pm 0.07$ &  2.20  \nl
147759.8&     17.6  & $ 0.34\pm 0.03$ &  6.38  \nl
147943.7&      5.1  & $ 0.30\pm 0.03$ &  1.65  \nl
        &     13.4  & $ 0.09\pm 0.03$ &  1.35  \nl
147979.7&      5.6  & $ 0.14\pm 0.02$ &  0.85  \nl
148249.2&      5.8  & $ 0.16\pm 0.04$ &  0.97  \nl
150155.3&     13.1  & $ 0.08\pm 0.03$ &  1.08  \nl
150186.7&      9.3  & $ 0.10\pm 0.04$ &  0.97  \nl
\enddata                              
\tablenotetext{a}{ All parameters were obtained from Gaussian fit. Note 
U147943.7 has two velocity components.}
 
\end{deluxetable}
\clearpage